\documentclass[letterpaper]{IEEEtran}
\usepackage{graphicx,times, amsmath, amsfonts, comment, color}
\usepackage{amssymb,epstopdf}
\usepackage[noend]{algorithmic}
\usepackage{algorithm}
\usepackage{multirow}

\newcommand{\beq}{\begin{equation}}
\newcommand{\eeq}{\end{equation}}

\begin{document}

\title{LTE/LTE-A Random Access for Massive Machine-Type Communications in Smart Cities}
\author{\IEEEauthorblockN{Md Shipon Ali, Ekram Hossain, and Dong In Kim} }
\maketitle

\begin{abstract}
Massive Machine-Type Communications (MTC) over  cellular networks is expected to be an integral part of  wireless ``Smart City''  applications. The Long Term Evolution (LTE)/LTE-Advanced (LTE-A) technology is a major candidate for  provisioning of MTC applications. However, due to the diverse characteristics of payload size, transmission periodicity, power efficiency, and quality of service (QoS) requirement, MTC poses huge challenges to LTE/LTE-A technologies. In particular, efficient management of massive random access is one of the most critical challenges. In case of massive random access attempts, the probability of preamble collision drastically increases, thus the performance of LTE/LTE-A random access  degrades sharply. In this context, this article reviews the current state-of-the-art proposals to control massive random access of MTC devices in LTE/LTE-A networks. The proposals are compared in terms of five major metrics, namely, access delay, access success rate, power efficiency, QoS guarantee, and the effect on Human-Type Communications (HTC). To this end, we propose a novel collision resolution random access model for massive MTC over LTE/LTE-A. Our proposed model basically resolves the preamble collisions instead of avoidance, and  targets to manage massive and bursty access attempts. Simulations of our proposed model show huge improvements in random access success rate compared to the standard slotted-Aloha-based models. The new model can also coexist with existing LTE/LTE-A Medium Access Control (MAC) protocol, and ensure high reliability and time-efficient network access. 
\end{abstract}


\section*{Introduction}
The term ``Smart City" represents an environment in which all the city assets are virtually connected and electronically managed. Smart utility, e-health, online school, e-library, online surveillance, environment monitoring, and connected vehicles are some of the smart city applications. For such an application, a huge number of autonomously operated, low-cost devices (i.e. sensor, actuator) need to be connected with physical objects. The communications between these autonomously operated devices are called Machine-Type Communications (MTC), and the MTC Devices (MTCDs) form an integral part of a smart city environment.  

Different technologies, such as wired networks, local area and short-range wireless networks, and cellular wireless networks, have been studied to enable massive MTC applications. However, due to the requirements of mobility, extended coverage area, security, diverse QoS, etc., a large percentage of MTCDs will  need to connect directly to cellular networks. 
The orthogonal frequency division multiple access (OFDMA)-based LTE\footnote{We use the term ``LTE" to refer to both ``LTE" and ``LTE-A" technologies.} technologies are major cellular technologies which will need to support the MTC applications in smart cities.

Random Access (RA) is the first step to initiate a data transfer using an LTE  network. According to 3GPP specifications, contention based RA occurs in following cases: (i) initial access to the network, (ii) recover Radio Resource Connection (RRC), and (iii) data transfer and location identifications  during RRC-connected state, when uplink is not synchronized. RA management is the most challenging task to support massive MTC in LTE systems. The MAC layer in LTE systems is based on the slotted-Aloha protocol, and severe congestion during random access is generally expected due to irregular and bursty access nature of transmissions by MTCDs. 

To resolve the RA congestion in LTE systems, different solutions have been proposed. These proposals mainly focus on five key performance metric: access delay, access success rate, QoS guarantee, energy efficiency, and the impact on HTC traffic. In this article, we provide a review of these proposals to solve the congestion problem during random access in terms of five aforementioned performance metrics. Some of the proposals are also discussed in 
\cite{3GPP2011}-\cite{Shirvanimoghaddam2015}. Nonetheless, most of the solutions are based on the collision avoidance technique, which simply restrict the arrival rate of access attempts. This results in large access delay, and therefore, the QoS requirements may not be satisfied for some MTCDs. This motivates us to develop a novel collision resolution based RA approach, where an $m-$ary contention tree splitting technique \cite{Jassen2000} is applied to resolve  collisions among preambles during random access. In this approach, the base station (BS), e.g. the evolved node B (eNB) in an LTE network resolves the random access collisions by scheduling the collided MTCDs into a set of reserved opportunities. In \cite{Madueno2014}, a different tree splitting RA model was studied. Different from that in \cite{Madueno2014}, our proposal is able to handle massive bursty traffic, and also can coexist with the existing LTE MAC protocol without any major modifications.

The rest of the article is organized as follows. We first review the contention-based RA process in LTE system. Major limitations of the existing approaches are then presented, where a particular MTC application is studied to understand the limitation of 
slotted-Aloha-based RA protocol. Next, we provide a survey of the existing RA congestion control proposals which is followed by our proposed collision resolution approach. Simulation results for the proposed approach are presented and compared with those for the standard LTE RA process. 

\section*{Contention-Based Random Access in LTE}
\subsection*{Random Access Preamble}
Random access preambles are the orthogonal bits sequences, called digital signature, used by UEs to initiate RA attempt. RA preambles are generated by cyclically shifting a root sequence, such that every preamble is orthogonal to each other. There are total $64$ preambles which are initially divided into two groups, i.e. contention-free RA preambles and contention-based RA preambles. The eNB reserves some preambles, say $N_{cf}$, for contention-free RA, and assigns distinct preambles to different UEs.  Rest of the preambles ($64-N_{cf}$) are used for contention-based RA, where each UE randomly generates one preamble \cite{3GPP2015}. 

\subsection*{Random Access Slot}
A random access slot (RA slot) refers to the LTE physical radio resources, called Physical Random Access CHannel (PRACH), in which RA preambles are mapped and transmitted to the eNB. In Frequency Division Duplex (FDD) operation, an RA slot consists of $6$ physical Resource Blocks (RBs) in frequency domain, while the time duration of each RA slot can be $1$, $2$, or $3$ subframe(s) depending on the preamble format \cite{3GPP2015}. There are a total of $864$ subcarriers in one RA slot which are equally distant at $1.25$ KHz. All $64$ preambles are mapped into $839$ RACH subcarriers, while the remaining $25$ subcarriers are used as guard frequency subcarriers\cite{3GPP2015}. 

In FDD operation, four different preamble formats are available based on preamble cyclic prefix duration ($T_{CP}$), and preamble sequence duration ($T_{SEQ}$) \cite{3GPP2015}. A UE can select an appropriate preamble under a specific format depending on the distance from eNB, maximum delay spread, amount of transmission resource needed to transmit RRC request, etc. On the other hand, the number of RA slots in each radio frame is defined by the preamble configuration index. For each preamble format $16$ different indices are available, where the eNB allocates radio resources as PRACHs. Depending on the system bandwidth, some LTE systems may not be able to use some preamble configuration indices. However, systems using $20$ MHz bandwidth are able to use all of the indices \cite{3GPP2015}. The eNB periodically broadcasts the preamble information as a part of System Information Block $2$ (SIB$2$) message. 
\begin{figure}[h]
\begin{center}
	\includegraphics[width=3.9 in]{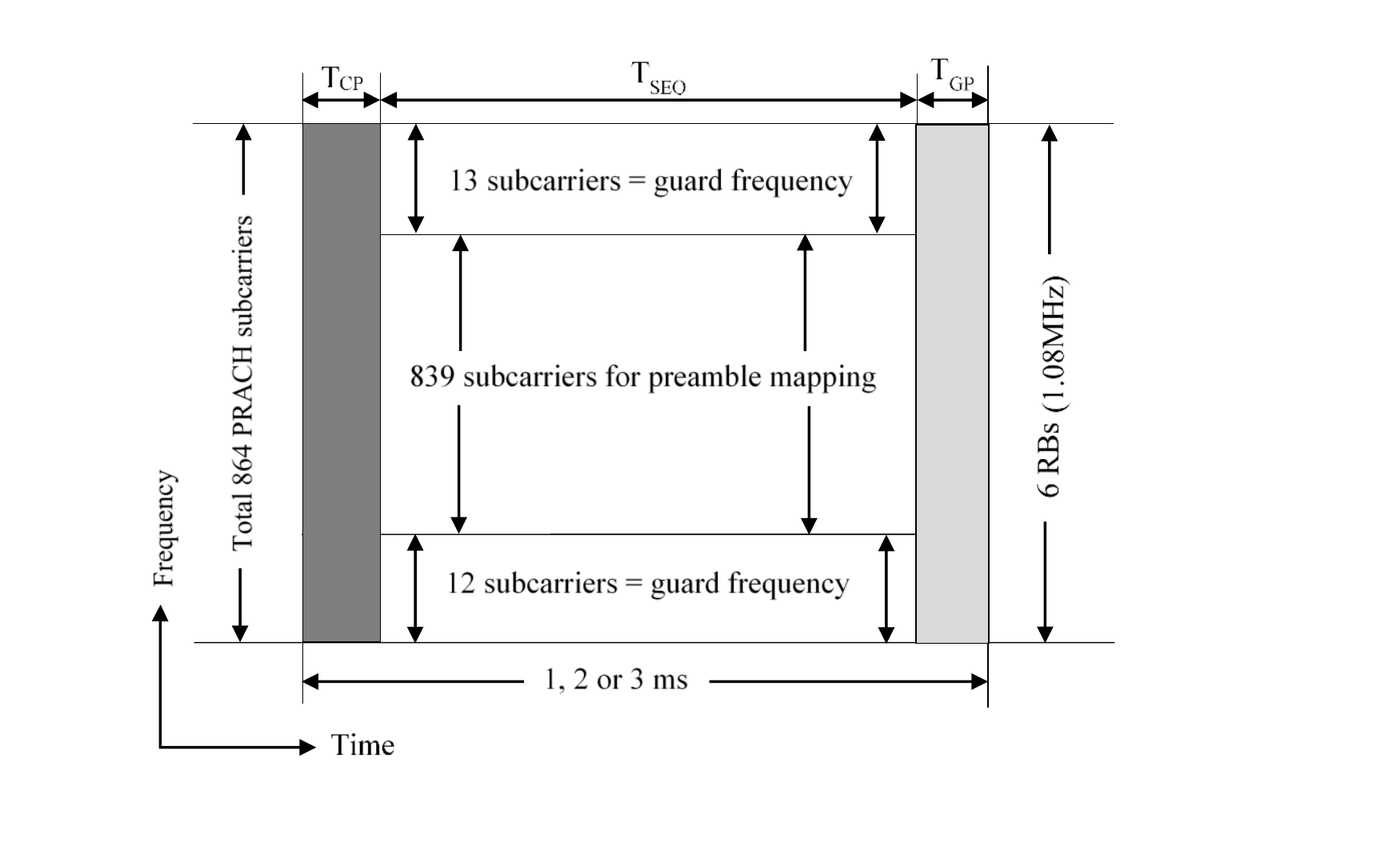}
	\caption{FDD-based RA slot in time-frequency resources.}
	\label{fig:alpha3}
 \end{center}
\end{figure}
\subsection*{Contention-Based Random Access Procedure}
When a UE is switched on or wakes up, it first synchronizes with the LTE downlink channels by decoding the Primary and Secondary Synchronization Signal (PSS \& SSS). The UE then decodes the Master Information Block (MIB), which contains information about the location of the downlink and uplink carrier configurations, thus gets the information of SIBs. All the RA parameters, i.e. RA slots, preamble formats, preamble configuration indices, etc. are contained in SIB$2$. Therefore, after decoding the SIB$2$, UEs can generate contention based RA attempt. The contention-based LTE RA procedure consists of four main steps as follows:
\begin{figure}[h]
\begin{center}
	\includegraphics[width=3.9 in]{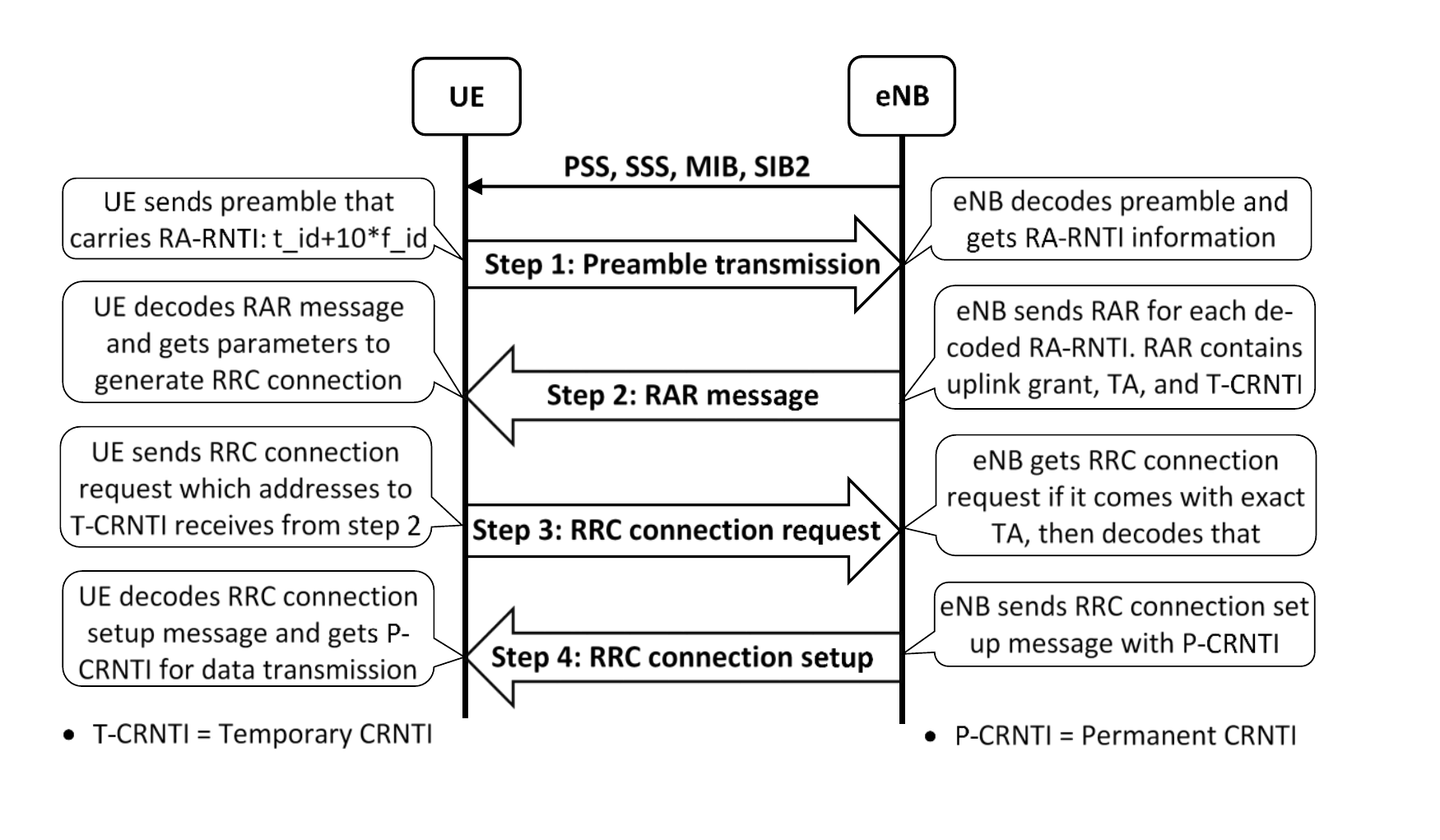}
	\caption{Contention-based RA procedure.}
	\label{fig:alpha3}
 \end{center}
\end{figure}
\subsubsection{Preamble transmission from UE to eNB}
In order to initiate a contention-based RA, a UE randomly generates one of the available contention-based preambles and sends that to the eNB at next available RACH slot. To select an appropriate preamble, the UE takes into account the current downlink path-loss, maximum delay spread, the size of transmission resource needed for RRC, and the required transmission power for RRC request. The eNB periodically broadcasts SIB$2$ message that suggests the UEs to choose appropriate preamble based on the aforementioned criteria. However, due to the orthogonal properties of RA preambles, the eNB can easily decode the different preambles unless multiple UEs transmit the same preamble at the same RA slot. The physical properties of RA preamble in PRACH contain the RA Radio Network Temporary Identity (RA-RNTI) and preamble configuration index information. After sending a preamble, the UE waits for a Random Access Response (RAR) window.

\subsubsection{Random access response from the eNB to UE}
After receiving the preambles on PRACH, the eNB calculates the power delay profile (PDP). If the estimated PDP is higher than a predefined threshold, the preamble is regarded as active. For each active preamble, the eNB decodes the RA-RNTI to find out the specific RA slot in which the preamble has been sent. After that, the eNB sends the RAR message to the decoded UEs on downlink control/shared channel. The RAR message contains a Timing Advance (TA) instruction to synchronize subsequent uplink transmissions, an uplink resource grant for RRC request, and a temporary Cell Radio Network Temporary Identifier (C-RNTI) which may or may not be made permanent at contention resolution time. The uplink grant part of RAR message contains 
all the necessary information to allocate resources for RRC attempt. However, if multiple UEs transmit the same preamble at the same RA slot, they will receive the same RAR message if the eNB could not detect the collision.

\subsubsection{RRC connection request from UE to eNB}
After receiving the bandwidth assignment at Step $2$, the UE sends an RRC connection request along with tracking area update and scheduling request. Step $3$ message is addressed to the temporary C-RNTI allocated in RAR message at Step $2$, and carries either a specific C-RNTI if the UE already has one (RRC-connected UEs), or an initial UE identity $-$ the Temporary Mobile Subscriber Identity (S-TMSI), or a random number. However, the colliding UEs, i.e. those  not detected at Step $2$, transmit RRC connection requests using the same uplink resources, thus another collision occurs at the eNB end.

\subsubsection{RRC connection setup from eNB to UE}
This step is called contention resolution stage. After decoding the RRC request message, the eNB acknowledges this to UE,  and sends RRC contention setup messages using the dedicated C-RNTI (if indicated in Step $3$ message), or  the temporary C-RNTI which also becomes dedicated from this stage. Successful UEs then acknowledge the eNB, and proceed for data transmission. However, the collided UEs, i.e. those which had sent RRC requests using the same uplink grant, will not receive feedback if their requests do not come with proper TA instruction. In this case, the collided UEs  will initiate a new RA access procedure after a maximum number of attempts for retransmission. 

\section*{Major Limitations of LTE Random Access}
In each RA slot, let us consider that $54$ preambles are utilized for contention-based random access, and each radio frame contains $2$ RA slots. Thus, the maximum number of RA opportunities per second are $10800$  $(=54\times2\times100)$, while simultaneous RA opportunities (preambles per RA slot) are still bounded by $108$. Also, if $30\%$ contention-based preambles are initially allocated for low data rate MTCDs, the maximum RA opportunity for low data rate MTCDs per second are $3240$. In addition, since LTE MAC protocol is slotted-Aloha based, the average RA success rate is around $37\%$. On the other hand, for massive MTC applications,  a single event can drive several thousands of MTCDs to access the network almost simultaneously, and consequently, huge preamble collisions are anticipated.
 
{\bf{\em An example scenario}}: Consider an earthquake monitoring scenario in a densely populated urban area. Assume that MTCDs are deployed in a cell of radius $2$ km with a density of  $60$ MTCDs per square kilometer. Thus, the density of  MTCDs  per cell is $754$ $(\approx \pi\times2^2\times60)$. Also, consider that the speed of seismic surface wave is $10$ km per second, which will result in $754$  access attempts by MTCDs in $200$ ms $(\frac{2\times1000}{10})$. In this case, the probability of preamble collision is around $30\%$ $ (\approx 1-e^{-\frac{754}{10800\times.2}})$ with $10800$ RA opportunities per second. However, if $30\%$ of the contention-based preambles are dedicated for low data rate MTCDs, then the probability of collision will be $69\%$ $(\approx 1-e^{-\frac{754}{3240\times.2}})$. 

Since the collision rate of slotted-Aloha system  increases exponentially with increasing rate of RA attempts, the random access in LTE networks is likely to be  unstable for massive MTC applications. 

\section*{Proposals to Improve LTE Random Access}
In this section, we review a wide range of RA congestion solution proposals in LTE systems.
The proposals are discussed under two classes, i.e. 3GPP specified solutions and non-3GPP specified solutions. Table I summarizes the proposals  in terms of five key performance metrics: access delay, success rate, energy efficiency, QoS guarantee, and impact on HTC.  

\begin{table*}
\begin{center}
\caption{Summery of various solutions of LTE Random Access Congestion}
\centering
\begin{tabular}{ |p{3.2cm}|p{3cm}|p{1cm}|p{1.1cm}|p{1.1cm}|p{1.15cm}|p{1cm}|p{1.1cm}|p{1.25cm}| }
\hline
\centering
\multirow{2}{*}{Proposal} & \centering \multirow{2}{*} {Sub-Proposal$/$Mechanism}	& \centering \multirow{2}{*}{Reference} & \centering Access	Delay & \centering Success Rate & \centering Energy Efficiency & \centering QoS   Assurance & \centering Impact on HTC & Performance Evaluation\\
\hline\hline
\multirow{3}{*}{Access class barring} & Individual ACB & $[5]$ & Varied & High & Medium & High & positive & No \\ \cline{2-9}
& Extended ACB & $[5],[7]$ & Varied & High & Medium & High & Positive & Yes \\ \cline{2-9}
& Cooperative ACB & $[8]$ & Varied & High & Medium & High & Positive & No \\ \cline{1-9}
\hline
MTC-specific backoff	& BI & $[5],[6]$ & High & Low & Low & No & Positive & Yes \\
\hline
Resource separation	& RACH split & $[5],[6]$ & High & Low & Low & No & Positive & Yes \\
\hline
Dynamic RACH allocation	& RACH add/drop & $[5], [6]$ & Medium & Medium & Medium & No & Negative & Yes \\
\hline
Slotted-access & Specific RA & $[5]$ & High & Very high & Very high & very low & Negative & No \\
\hline
\multirow{3}{*}{Pull-based access} & Individual paging & $[5]$ & Medium & Medium & Medium & No & Negative & No \\  \cline{2-9}
& Group paging & $[5],[11]$ & Medium & Medium & Medium & No & Negative & No \\  \cline{2-9}
& Group access & $[11]$ & Low & High & Very high & No & Positive & Yes \\  \cline{1-9}
\hline
Self-optimization & ACB, RA split \& add/drop & $[9]$ & Low & High & High & High & Positive & No \\
\hline
Prioritized access & ACB, RACH split & $[10]$ & Varied & Medium & Medium & High & Positive & No \\
\hline
Code-expanded & Code wise access &  $[12]$ & Low & High & Very low & No & \centering -- & Yes \\
\hline
Spatial grouping & Preamble reuse & $[13]$ & Low & High & Medium & No & \centering -- & Yes \\
\hline
Guaranteed access & Instant control & $[14]$ & Low & High & Low & High & Positive & No \\
\hline
Non-Aloha based RA & Analog fountain code & $[15]$ & Low & High & Low & No & \centering -- & No \\
\hline
\end{tabular}
\end{center}
\end{table*}
\subsection*{3GPP Specified Solutions}
In \cite{3GPP2011}, 3GPP specified the following six distinct solutions of LTE RA congestion due to massive MTC applications. 
\subsubsection{Access class barring}
Access Class Barring (ACB) is a well-known tool to control the RA congestion by reducing the access arrival rate. ACB operates on two factors: a set of barring Access Classes (ACs) in which devices are classified, and a barring time duration ($T_b$). Depending on the RA congestion level, the eNB broadcasts an access probability $p$, and barring time duration $T_b$ as a part of SIB$2$. The UEs,  which intend to access the network randomly, generate their own access probability $q$ according to the  AC they belong to. If $q \leq p$, the UE gets permission to access the network, otherwise it is barred for an ACB window $T_b$. To support massive MTC along with HTC, the 3GPP allowed separate AC(s) for MTCDs \cite{3GPP2011}. Also, depending on the QoS requirements, the MTCDs are sub-grouped into different ACs. The ACB model has further been optimized in terms of efficient AC management and dynamic updating of ACB parameters. The 3GPP also specified two different ACB mechanisms for massive MTC over LTE system as follows:
\begin{enumerate}
\item[\textbf{i}] \textbf{Individual access class barring$:$} Individual ACB is mainly proposed to achieve better QoS requirements. Each individual device or a group of devices having same QoS requirements are classed together. However, in massive MTC, tens of thousands MTCDs are expected to serve under a single cell, where many of them have distinct QoS requirements. Thus, individual ACB is not efficient 
for massive MTC. 

\item[\textbf{ii}] \textbf{Extended access class barring$:$} Extended Access Barring (EAB) was proposed as the baseline solution to relieve the RA congestion for massive MTC in LTE systems \cite{3GPP2011}, \cite{3GPP2010}. In EAB, the MTCDs are classified based on their QoS requirements, and the EAB algorithm dynamically barres and unbarres the low-priority MTCDs depending on the RA arrival rate. Thus, EAB ensures timely network access to the delay-constrained MTCDs. To enhance the performance of EAB, several algorithms were proposed. In \cite{3GPP2012}, 3GPP RAN-Group proposed an improved EAB algorithm which  optimizes the EAB parameters depending on the congestion coefficient, i.e. the ratio of collided preambles and successful preambles over a certain time duration.
\end{enumerate}

Apart from 3GPP specified improvements, the authors in \cite{Lien2012} proposed cooperative ACB which is mainly targeted for heterogeneous networks (HetNets). In this scheme, the cooperative eNBs jointly determine their ACB parameters, thus the RA congestion is distributed among the cooperating eNBs.

Although the ACB schemes provide a better RA success rate, the access probability $p$ might need to be set to a very restrictive value in case of high congestion scenarios. Therefore, long access delay will badly impact the low-priority applications. In addition, the ACB models are not capable of handling RA congestion due to the massive bursty access attempts by MTCDs.

\subsubsection{MTC-specific backoff}
Backoff mechanism is a common solution to control random access in cellular networks. The basic idea behind this backoff mechanism is, it discourages the UEs to seek the access opportunity for a time duration, called Backoff Interval (BI), if their first attempt failed due to collision or channel fading. If a device  fails  second time to get access, it will be subjected to a larger BI than the previous one. In MTC-specific backoff, MTCDs are subjected to a larger BI compared to the HTCDs. 

\subsubsection{Dynamic resource allocation}
Dynamic allocation of RACH is a straightforward solution for the RA congestion problem. Under this scheme, the eNB can increase the RACH resources in frequency domain, time domain, or both, based on the level of RA congestion \cite{3GPP2011}. In time domain, the eNB can allocate up to ten subframes as PRACH. In frequency domain, additional $1.08$ MHz of bandwidth can be allocated for PRACH. It is worth noting that, if more uplink resources are utilized as PRACH, there might be shortage of data channel. In \cite{3GPP2010}, 3GPP evaluated the performance of dynamic RACH allocation scheme, and recommended it as the primary solution to the RA congestion problem for massive MTC.

\subsubsection{Slotted random access}
In slotted RA scheme, each MTCD is assigned to a dedicated RA opportunity and only allowed to perform RA in its own dedicated access slot \cite{3GPP2011}. All the access slots comprise an RA cycle, and the eNB periodically broadcasts the parameters of the RA cycle and access slots. For a large number of MTCDs, the duration of the RA cycle is likely to be very large, thus MTCDs might experience long access delay. 
In addition, since each LTE RA slot consists of $64$ RA opportunities, there is a strong possibility that all 64 access attempts are made within a single RA slot, thereby giving rise to collision in a slotted-Aloha based MAC system. Moreover, in slotted RA scheme, while there could be very high load in some slots, some other slots may remain underutilized. 

\subsubsection{Separate random access resources}
To save HTC devices (HTCDs) from RA congestion, separate RACH for MTCDs has been proposed. The separation of resources can be made either by allocating separate RA slots for HTCDs and MTCDs, or by splitting the available preambles into HTC and MTC subsets \cite{3GPP2011}. To ensure QoS guarantee for HTC, some studies proposed to utilize full resources by HTCDs, whereas MTCDs are restricted to their own subset. Although the RACH separation scheme potentially reduces the negative impact on HTCDs, MTCDs might experience serious congestion because the available resources are  reduced for MTCDs, and the performance tends to be worse under high M2M traffic load.

\subsubsection{Pull-based random access}
All of the above RA congestion solutions use a push-based approach, where the RA attempts are performed arbitrarily by individual devices. Pull-based RA model \cite{3GPP2011} is an alternative approach where the devices are only allowed to perform RA attempt when they receive any paging message from the eNB. The pull-based RA model is suitable where the MTCDs  transmit information to their server in an on-demand basis. Under this scheme, a server requests an eNB to send paging message to the respective MTCDs in order to report their outputs. Therefore, it is a centralized approach in which the eNB can completely control the RA congestion by delaying the paging message. Some MTCDs usually transmit data to their servers in a periodic manner without any request from the server. Thus, by introducing paging for these periodic applications, the eNB can also control RA congestion. However, for massive MTC, it is a challenging task to page a large number of devices which requires extra control channels. 

To reduce the paging load, the 3GPP proposed a group paging method in \cite{3GPP2011}. Group paging enables paging a large number of MTCDs in one paging occasion, thus it reduces the usages of paging channels. However, all the MTCDs included under a group paging simultaneously perform the RA attempts. Therefore, the number of MTCDs included under a group paging is bounded by the RACH resources.

\subsection*{Non-3GPP Random Access Solutions}
Besides the 3GPP specified solutions, different academia, industry, and government bodies also proposed several solutions of RA congestion in order to support massive MTC over LTE networks. Some of the proposals contain distinct characteristics, while others showed improved performances of 3GPP specified solutions. Important proposals are studied in following subsections.
\subsubsection{Self-optimization overload control random access}
Self-optimization overload control (SOOC) approach combines RA resource separation, dynamic RA resource allocation, and dynamic access barring scheme \cite{Lo2011}. This approach was proposed for MTC applications, which uses RA resources separate from those for HTC. Also, the MTCDs are sub-grouped into two units: 
high-priority MTCDs and low-priority MTCDs. Under this model, MTCDs send RRC requests along with a counter value which indicates the number of RA attempts they had done before receiving the successful RAR message. By observing the counter value, the eNB estimates the RA congestion level. Depending on the congestion level and available uplink radio resources, the eNB either increases the RA resources, or decreases the access probability of low-priority MTCDs, or takes both actions together. Finally, the new parameters of RA resources and access probability are broadcast as a part of SIB$2$ message. 

\subsubsection{Prioritized random access}
Prioritized RA is another optimization approach based on RA resource separation and ACB mechanism. Under this approach, applications are divided into five classes: HTC, high-priority MTC, low-priority MTC, scheduled MTC, and emergency service \cite{Cheng2011}. The available RACHs are virtually separated into three groups: HTC, random MTC, and scheduled MTC \& emergency service \cite{Cheng2011}. A prioritized access algorithm is developed to ensure QoS guarantee for the application classes as well as virtual groups. The prioritization is achieved by introducing distinct backoff window sizes for different classes. 

\subsubsection{Group-based random access}
Group-based RA approach is an extension of  pull-based group paging RA model. Under this scheme, MTCDs under a group paging occasion form one or more access group(s). MTCDs of an access group are assigned an identity, called a group index, corresponding to their access group ID and paging group ID. Formation of access groups can be based on different criteria, i.e. belonging to the same server, having same specifications, similar QoS requirements, located in a specific region, etc. However, the key aspect enabling the group access mechanism is that all group members are in close proximity of each other such that TA estimation for the group delegate is valid for all group members \cite{Farhadi2013}. In group-based RA process, a single preamble is used for all MTCDs of each access group, but only the group delegate is  responsible for communicating with the eNB. The eNB selects the group delegate based on different metrics such as channel condition, transmission power, etc. A bearer dedicated to each access group is created following the standard process when the group delegate connects to the network, but it is transparent to all group members. 
However, the group members need to share their data with the group delegate.

\subsubsection{Code-expanded random access}
In code-expanded RA model, an RA attempt is initiated by sending a set of preamble(s), called RA codeword, over a predefined number of RA slots, instead of sending simply a single preamble at any arbitrary RA slot. In this method, a virtual RA frame is considered which consists of a group of RA slots, or a set of preambles in each RA slot. MTCDs need to send multiple preambles over each virtual RA frame, thus making a codeword. At the receiver end, the eNB identifies the individual RA attempts based on the identical codeword perceived inside it. Let us consider that $L$ RA slots are formed into a virtual RA frame and each RA slot consists of $M$ preambles, thus the available number of codewords is $[((M+1)^L)-1]$ \cite{Pratas2012}, which increases the RA opportunities significantly without increasing any physical resources. 

\subsubsection{Spatial-group-based reusable preamble allocation}
The main idea behind this RA model is to spatially partition the cell coverage area into a number of spatial group regions, and the MTCDs in two different spatial group regions can use the same preambles at the same RA slot if their minimum distance is larger than the multi-path delay spread. It is possible due to the fact that the eNB is able to detect simultaneous transmission of identical preambles from different nodes if the distance between the detected picks is larger than the delay spread. In the RAR message, the eNB sends distinct RAR for each of the detected nodes, but all the RAR are addressed to the same preamble since the nodes had sent the RA request using the same preamble. However, each RAR contains different TA values for different devices, and the devices can detect the correct RAR by matching their estimated TA with the set of TAs in the RAR message. The authors in \cite{Jang2014} obtained simulation results assuming a cell radius of $2$ km, where the minimum distance between the identical usable preamble groups are $0.2$ km. However, although a small delay spread provides more RA opportunities, it can also cause  mis-detection of preambles. 

\subsubsection{Reliability guaranteed random access}
Generally, RA congestion is detected upon preamble collision rate, and the control schemes deal with high RA load by optimizing the control parameters. The updates are periodically broadcast on SIB$2$, where the cycle of SIB$2$ takes up to $5$ seconds \cite{3GPP2015}. Therefore, in case of bursty traffic, it is not possible to update the parameters immediately to maintain the QoS requirements of MTCDs. To address this issue, the authors in \cite{Madueno2015} proposed a proactive approach where the load estimation is performed within one RA slot and the parameters are optimized instantly. Under this model, a number of RA slots, say $L$, are formed into an RA frame which consists of two phases: a load estimation phase (one slot), and a serving phase containing $L-1$ RA slots. MTCDs are also sub-grouped according to their QoS requirements, and each sub-group is assigned different preambles in the estimation phase. However, all MTCDs need to perform RA attempts during the estimation phase. Based on the rate of preamble collision, the eNB estimates the number of contention users in each group. The eNB then distributes the $L-1$ RA slots in the serving phase according to the QoS requirements.  After that, the MTCDs again send their RA requests in their specific RA slots. 

\subsubsection{Non-Aloha-based random access}
The limited number of RA preambles is the main bottleneck of slotted-Aloha-based RA for massive MTC in LTE networks. Recently, the authors in \cite{Shirvanimoghaddam2015} proposed a RA model based on the capacity-approaching Analog Fountain Code (AFC). AFC-based RA  combines  multiple access with resource allocation. In this model, multiple MTCDs can send RA requests by using the same preamble, and then data transmission also occurs within the same RB. The available contention-based RA preambles are sub-grouped based on the QoS requirements of MTCDs. The RA process has two phases, i.e. contention phase and data transmission phase. In the contention phase, all the MTCDs with the same QoS are grouped together and 
initiate RA attempt by using predefined preamble(s). Depending on the received preamble power, the eNB estimates the number of contended MTCDs per preamble, and broadcasts this information to all contending MTCDs for each of the detected preambles. The MTCDs which sent the same preamble obtain the information about total number of candidate MTCDs for that preamble. Based on the number of collided MTCDs, each MTCD generates an orthogonal random seed and shares it with the eNB. Therefore, both the eNB and MTCD can construct the same bipartite graph to perform AFC encoding and decoding for subsequent communications \cite{Shirvanimoghaddam2015}. 

\section*{Collision Resolution-Based  Random Access}
\subsection*{Model}
The basic idea behind the collision resolution based RA (CRB-RA) model is to ensure RA reattempts from a reserved set of preambles if the current attempt is detected as a collision. The number of preambles in each reserved set is optimized according to the rate of collision at each level. In this model, separate RA preambles are used for HTC and MTC, where the collision resolution technique is only applicable for MTC. A number of RA slots form a virtual RA frame, and the eNB broadcasts SIB$2$ at the end of each virtual RA frame. The eNB can allocate new RACH resources into the virtual RA frame if the collision rate is increased at certain thresholds, thus the size of virtual RA frame is optimized depending on the rate of preamble collision. Meanwhile, the duration of each virtual RA frame is adjusted upon the QoS requirements of high-priority MTCDs. In addition, each contending UE (MTCD/HTCD) transmits its identity, UE-ID, along with randomly generated preamble for RA attempt \cite{Zhang2015}. Some RACH subcarriers are used to map UE-ID such that the UE-IDs of different preambles are orthogonal to each other \cite{Zhang2015}. However, if multiple UEs transmit the same preamble at the same RA slot, the eNB is unable to decode their UE-IDs, and thus considers it as a collision. For each collided preamble, the eNB assigns a set of new preambles (say $m$) to the collided UEs if the collided preamble belongs to the MTCDs. In the RAR message, the eNB instructs the collided MTCDs to retransmit on the reserved preamble set in the next available virtual RA frame. On the other hand, if the collided preamble arrives from HTCDs, the eNB does not send any RAR feedback, thus the collided HTCDs re-initiate a new RA procedure at the next available RA slot.  
 
In the next virtual RA frame, the collided MTCDs retransmit the RA requests using preambles from the reserved set, while the others are not allowed to use that set. The eNB imposes restriction by broadcasting the information as a part of SIB$2$. If the collided MTCDs collide again within the preassigned $m$ preambles, then another new set of preambles will be allocated accordingly. This process will continue until the eNB properly decodes each preamble with individual UE-ID. Therefore, an optimistic $m-$ary splitting tree algorithm is developed for each collision. However, based on the collision rate, the eNB can also utilize dynamic ACB mechanisms to facilitate channel access for high-priority MTCDs.  

\begin{figure}[h]
\begin{center}
	\includegraphics[width=3.8 in]{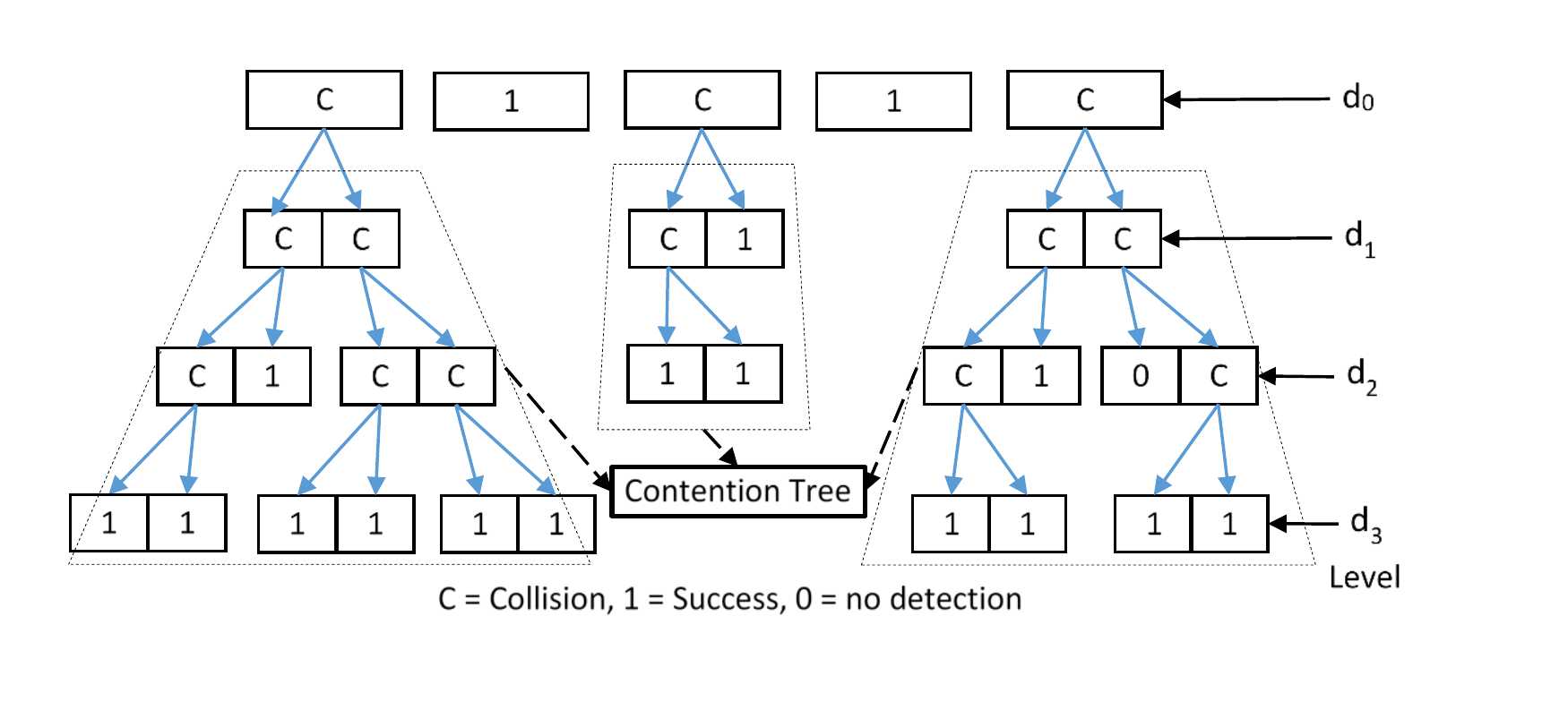}
	\caption{Illustration of collision resolution RA model.}
	\label{fig:alpha3}
 \end{center}
\end{figure}

Fig. 3 illustrates our proposed CRB-RA model by using a binary ($m=2$) splitting tree algorithm. In this model, the basic splitting tree algorithm is slightly modified to resolve the RA problem in LTE. The root of the new model, where collisions initially occur, consists of all number of contention-based preambles (say $q$) of a virtual RA frame. Let us denote the root as level $0$. For each of single collision at level $0$, a new set of $m$ preambles is reserved at level $1$. Similarly, $m$ preambles are also reserved at level $2$ for each collision detected at level $1$, and the process continues until  the collision is resolved. Therefore, an $m$-ary tree is developed for every preamble collision detected at level $0$, but the root of each individual tree is at level $1$. 

In the CRB-RA model, the number of preambles in each reserved set ($m$) is dynamically adjusted according to the collision rate. Also, each level of contention tree is resolved at individual virtual RA frame. In a particular virtual RA frame, if the collision rate is sufficiently high, then more  preambles are required for the reserved set  where the value of $m$ would also be high. For example, in case of full collision, the maximum number of preambles required at any level is $(m^d \times q)$, where $d$ indicates the level of the tree. However, if the value of $m$ is set to a high value, then each level of contention tree requires more time to resolve. On the other hand, if the value of $m$ is set to a low value, the number of level would be high. Therefore, the access delay of CRB-RA mainly depends on the proper selection of $m$. The general algorithm of our proposed CRB-RA model for two different collision thresholds is presented as follows:
 \\
\rule[.1ex]{\linewidth}{1.5pt}
\textbf{Algorithm: Collision resolution based random access.}
\rule[.8ex]{\linewidth}{1pt}
\begin{enumerate}
\item[\bf{1.}] Set collision threshold: $x$, $y$; $x < y$
\item[\bf{2.}] Set preambles per contention tree slot: $m_0$\footnote{Initial value of $m$; for optimal resource utilization, $m_0 = 3$.}, $m_x$, $m_y$; $m_0 < m_x < m_y$ 
\item[\bf{3.}] Set additional RA slot: $\Delta_x$, $\Delta_y$; $\Delta_x \leqslant \Delta_y$
\item[\bf{4.}] Check preamble collision rate: $\kappa$
\item[\bf{5.}] While $\kappa \neq 0$ 
\item[\bf{6.}] If $y > \kappa \geqslant x$, then set $m = m_x$ and $\Delta = \Delta_x$
\item[\bf{7.}] Elseif  $y  \leqslant \kappa > x$, then set $m = m_y$ and $\Delta = \Delta_y$
\item[\bf{8.}] Else $m = m_0$, and $\Delta$ unchanged
\item[\bf{9.}] Reserve $m$ preambles for each collision 
\item[\bf{10.}] Send RAR to collided MTCDs for reattempt RA
\item[\bf{11.}] Broadcast the updated RA resources on SIB2.
\end{enumerate}
\rule[.2ex]{\linewidth}{1.5 pt}
\subsection*{Performance Analysis}
We evaluate the performance of our proposed CRB-RA model in terms of the average number of preamble retransmission and average outage probability. The results are compared with standard slotted-Aloha-based RA model. The energy efficiency and access delay of proposed CRB-RA model are also discussed based on the outage probability and average number of preamble retransmissions. It is assumed that massive access requests are attempted, i.e. as in the earthquake monitoring scenario discussed before. Each preamble can be successfully detected (collision/active/ideal) at the eNB end. Also, since collisions are detected before RAR feedback, the BI technique is not applied here. To simplify the simulations, issues such as mis-detection, propagation delay, and device processing time are not considered. In addition, we simulate our proposed CRB-RA model by considering a fixed contention slot size ($m$ is fixed). All the simulations are done based on the 3GPP parameters  \cite{3GPP2015}, where the initial PRACH configuration index is $6$ ($2$ RA slots per radio frame) and the maximum RA retransmission limit is $10$. In each initial RA slot, $30$ contention-based preambles are used for MTC. Also, depending on the collision rate, the eNB allocates up to $10$ RA slots per radio frame.
\begin{figure}[h]
\begin{center}
	\includegraphics[width=3.6 in]{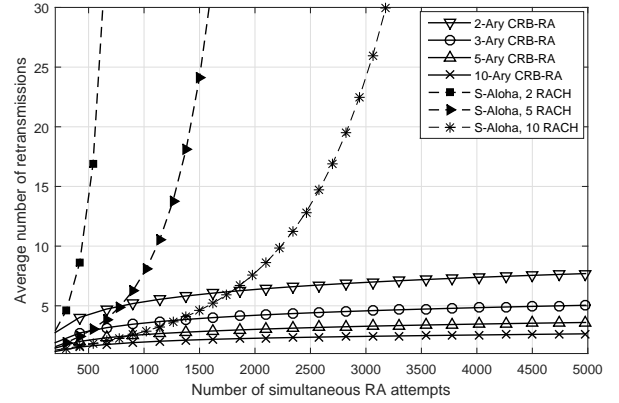}
	\caption{Average number of retransmissions for the proposed RA model and the slotted-Aloha-based RA model.}
	\label{fig:alpha3}
 \end{center}
\end{figure}
\begin{figure}[h]
\begin{center}
	\includegraphics[width=3.6 in]{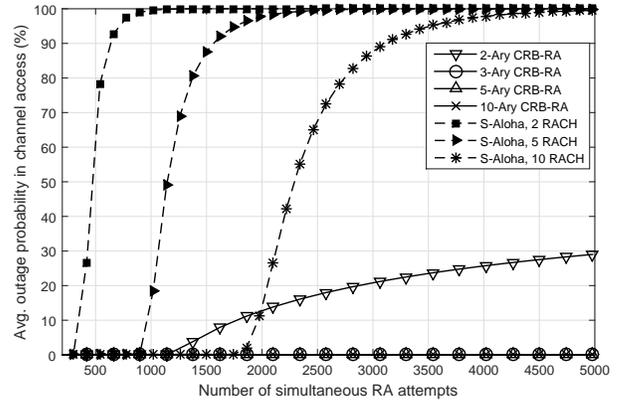}
	\caption{Average outage probability in channel access vs. number of simultaneous RA attempts.}
	\label{fig:max_error_plot}
 \end{center}
\end{figure}

Fig. 4 shows the average number of RA attempts required to successfully decode each MTCD with respect to the number of simultaneous\footnote{We use the term ``Simultaneous RA attempts" to refer to the number of RA attempts within one radio frame.} RA attempts. It is apparent that, for any arbitrary RA attempt, the proposed CRB-RA model ensures network access within a limited number of retransmissions, while a large number of retransmissions are required in case of standard slotted-Aloha-based RA model. The slotted-Aloha-based RA scheme with peak preamble configuration index ($10$ RA slots per radio frame) needs on average more than $30$ retransmission attempts for one successful access, when the number of simultaneous RA attempts is $3200$ or higher. It is noted that, in Fig. 4, the CRB-RA model utilizes only two RA slots in each radio frame.

Also, Fig. 5 shows the average RA outage probability of MTCDs as a function of the number of simultaneous access attempts. It is evident that, by setting an appropriate number of preambles (value of $m$) per contention slot, the CRB-RA model can reduce the outage in network access significantly. The standard slotted-Aloha-based RA system with $2$ RA slots per radio frame shows an average outage rate of around $70\%$ for $500$ simultaneous RA attempts. In addition, with the maximum number of RA slots per radio frame, the standard slotted-Aloha-based RA system shows around $70\%$ average outage probability for $2500$ RA attempts per radio frame. Therefore, massive multiple access by MTCDs will make the system unstable. However, in contrast, with minimal preambles per contention slot ($m=2$), although the proposed CRB-RA model may result in a non-zero outage probability for a large number of simultaneous RA attempts, by optimizing the slot length ($m$), the outage probability in channel access can be made very small. 

In addition, since the average number of RA retransmission requirement in CRB-RA model is very low compared to that for slotted-Aloha based RA, the proposed RA model is very efficient for power-constrained MTC applications. For the same reason, the access  delay for the CRB-RA model would also be much lower in comparison to that for slotted-Aloha-based RA models.

\section*{Conclusion}
We have outlined the major challenges of existing LTE MAC layer to bring massive ``smart city" applications into mass market. We have also  reviewed a wide range of LTE MAC layer congestion control proposals from the perspective of massive MTC applications. Some proposals can potentially handle high rate of RA requests, but the solutions are not capable of managing massive bursty access attempts. To solve this problem, we have proposed a novel collision resolution-based RA model, which can effectively manage massive RA requests. Also, our proposed RA method can coexist with existing LTE MAC protocol without any modification. Simulation results  have shown that the collision resolution RA model provides  reliable and time-efficient access performance.
Although we have simulated our model with a fixed size of reserved preamble set, the proposed model exhibits a multi-dimensional optimization problem, where the number of preambles per contention tree slot, and the size and duration of virtual RA frame can be optimized based on the preamble collision rate, available radio resources, and delay constraints. 

\bibliographystyle{IEEE}

\end{document}